# Selective effects of selenium substitution on superconductivity and magnetic memory in 4Hb-$TaS_2$

Shalev Gur [1], Sajilesh K.P.[2] , Shubam Purwar[2], Irena Feldman [2] ,Ilay Mangel[2], Nitzan Ragoler[2], Amit Kanigel [2*] and Beena Kalisky [1*]

1. Bar Ilan University, Ramat Gan, Israel
2. Technion, Haifa, Israel

*amitk@physics.technion.ac.il, *beena@biu.ac.il

**Disorder plays a central role in correlated quantum materials, but its effects can be highly selective when several electronic orders coexist. The van der Waals heterostructure 4Hb-$TaS_2$ provides such a case, combining superconductivity, charge-order tendencies, strong scattering, and magnetic memory manifested through spontaneous vortices. Here we use Se substitution as a controlled local perturbation of the disorder landscape in 4Hb-$TaS_2$. We find that superconductivity remains remarkably robust while the magnetic memory is strongly suppressed at higher Se concentrations. These results show that Se substitution does not create the hidden magnetic memory, but instead selectively weakens its macroscopic manifestation. More broadly, they reveal a sharp separation between robust superconducting and hidden-memory phases, and a highly disorder-sensitive spontaneous-vortex response in 4Hb-$TaS_2$.**

4Hb-$TaS_2$ (hereafter 4Hb) is a van der Waals heterostructure composed of alternating layers of Mott-insulating 1T-$TaS_2$ and superconducting 2H-$TaS_2$ [1–4]. This hybrid stacking gives rise to a type-II unconventional superconducting phase and a time-reversal symmetry-breaking (TRSB) state below Tc=2.7 K, consistent with a multicomponent superconducting order parameter [5–7].

Another unusual property of 4Hb is a magnetic memory effect that can be trained in the normal state and subsequently manifests below Tc through the appearance of spontaneous vortices [8]. The memory survives up to a characteristic temperature (T* ~ 3.6K, above Tc, despite the absence of a detectable magnetic signal in this temperature range. This unusual coexistence of magnetic memory and superconductivity has motivated several theoretical proposals involving Kondo physics, emergent correlated states, and unconventional superconductivity [9–13].

Se substitution provides a useful way to perturb 4Hb-$TaS_2$ without changing its basic layered structure. Because Se replaces S isovalently on the chalcogen site, it is expected to primarily modify the local chemical and structural environment rather than act as a conventional carrier-doping parameter. Recent STM/STS measurements on samples spanning the same Se-substitution series, including nominally Se-free crystals, revealed an intrinsically inhomogeneous electronic landscape associated with the incommensurability between the alternating 1T and 1H layers [14]. This incommensurability locally modulates the interlayer spacing, hybridization, and charge transfer between the correlated 1T and metallic 1H layers, linking structural inhomogeneity to charge-density-wave order, electronic correlations, and superconductivity [14]. In parallel, Se/S substitution has been proposed to influence superconductivity through disorder-induced delocalization in the 1T-derived electronic sector [15]. These results highlight disorder and structural inhomogeneity as potentially important ingredients in the electronic properties of 4Hb-$TaS_2$.

The sensitivity of this material family to disorder is reflected in reports of a correlation between chemical substitution and the suppression of charge-density-wave (CDW) order in 4Hb-$TaS_{2-x}Se_x$ and 2H-$TaSe_{2-x}$ [16,17], as well as superconductivity in 1T-$TaS_{2-x}Se_x$ [18–20]. Similarly, in monolayer 2H-$TaS_2$, reducing the thickness to the two-dimensional limit is accompanied by a suppression of the CDW order and a concomitant enhancement of Tc, suggesting a relationship between CDW order, superconductivity, and disorder [21].

More broadly, these observations place 4Hb within a wider class of correlated quantum materials in which disorder can play a deeply ambivalent role. In ferromagnetic systems, disorder can enhance magnetic ordering by disrupting competing antiferromagnetic interactions, leading to increased transition temperatures and modified thermodynamic responses; in diluted magnetic semiconductors, positional disorder has been shown to strengthen ferromagnetic correlations rather than suppress them [22,23].

The superconducting order parameter and the microscopic nature of the scattering. For a conventional isotropic s-wave state, non-magnetic disorder is largely innocuous, as expressed by Anderson's theorem [24], whereas for sign-changing order parameters impurity scattering can in principle be pair breaking [25]. Experimentally, however, a wide range of behavior is observed even among unconventional superconductors. In $Sr_2RuO_4$, Tc is rapidly suppressed by non-magnetic disorder [26], while in multiband iron-based superconductors the suppression rate depends strongly on the character of the scattering and can remain consistent with sign-changing s± pairing [27]. Even more strikingly, unconventional superconductivity in Sn-substituted CeRhIn5 survives despite a very large increase in residual impurity scattering [28]. These examples emphasize that substantial normal-state scattering is not by itself incompatible with unconventional superconductivity, and that the response to disorder must be considered in the context of the electronic and gap structure of each material.

Here, we use dilute Se substitution as a controlled local perturbation and compare its effect on several properties of 4Hb-$TaS_2$. We find that all superconducting samples in our series, including nominally Se-free crystals, exhibit low residual-resistance ratios, indicating substantial scattering even without intentional Se substitution. Moreover, the RRR remains comparable across the Se-substitution series, while the superconducting transition temperature changes only weakly. In striking contrast, the spontaneous-vortex manifestation of the magnetic memory is rapidly suppressed with increasing Se concentration. Notably, the original observation of magnetic memory was made in crystals containing a small amount of Se [8], whereas we find an even stronger memory response in Se-free 4Hb-$TaS_2$, demonstrating that Se substitution is not required for the effect. Thus, rather than providing a simple tuning from a clean to a disordered superconductor, dilute Se substitution reveals a highly selective response of the different electronic phenomena in 4Hb-$TaS_2$.

Single crystals of 4Hb-$TaS_{2-x}Se_x$ were grown using the chemical vapor transport (CVT) method. A stoichiometric mixture of tantalum (Ta), sulfur (S) and selenium (Se) was sealed in a quartz ampoule under vacuum. The mixture underwent a solid-state reaction, forming a boule of 4Hb-$TaS_{2-x}Se_x$. The boule was crushed and placed in a 200-mm-long quartz ampoule with a diameter of 16-mm. Iodine was added as a transport agent, and the ampoule was sealed under vacuum. The ampoule was then placed in a three-zone furnace, where the hot ends were heated to 800$^o$ C, while the middle section was kept at 750$^o$ C. After about 30 days, the ampoule was quenched in cold water. Crystals a few mm in size were found in the cold part of the ampoule.

The crystal structure and chemical composition were verified using X-ray diffraction and electron energy-dispersive spectroscopy in a scanning electron microscope. The measurements show that the actual amount of selenium in the crystals is slightly lower than the amount of selenium in the starting mixture (see supplementary material Fig.S1 for details [29]). For convenience, we use the nominal Se concentration values throughout the paper.

In Supplementary Fig.2 we show the residual-resistance ratio (RRR) and Tc as a function of Se concentration [29]. The RRR remains low and changes only weakly across the substitution series, indicating that all samples, including the nominally Se-free crystal, are in a strongly scattering regime. Despite this, the superconducting transition temperature remains remarkably robust, decreasing by no more than 10% even in samples with up to 20% Se substitution. This robustness of Tc is notable given prior evidence for a non-(s)-wave superconducting order parameter in this system [5,6].

Next, we examine the effect of disorder on the hidden memory. We carried out magnetometry and susceptometry measurements using scanning superconducting quantum interference device (SQUID) microscopy [30,31]. With a micron-sized pickup loop surrounded by a one-turn coil for applying a local AC magnetic field, the SQUID can simultaneously measure both DC and AC magnetic flux, the latter resulting from diamagnetic screening currents canceling the field generated by the field coil [32,33]. We used scanning SQUID to capture maps of the local magnetic flux, mapping the configuration of vortices in the superconductor.

In conventional type-II superconductors, magnetic flux can penetrate the material in the form of quantized vortices, but only in the presence of an external magnetic field. 4Hb, however, exhibited the emergence of vortices even in lack of a magnetic field. These spontaneous vortices were shown to be trainable by cooling through $T^* = 3.6K$ with an external magnetic field. $T_c < T < T^*$ is a normal non-magnetic state, without any detectable magnetic signals (down to scanning SQUID sensitivity), giving it the name of the hidden magnetic memory effect. To probe the hidden magnetic memory phase, we implemented a series of cooling, heating, and re-cooling protocols under various magnetic fields while imaging the magnetic landscape to quantify the vortex density. To identify the magnetic memory we search for vortices appearing after cooling in zero-field. Such spontaneous vortices appear only after training the sample by cooling it through T* in the presence of a magnetic field, which is why this field is referred to as the training field. The number of spontaneous vortices indicates the strength of the memory and depends on the training field [8].

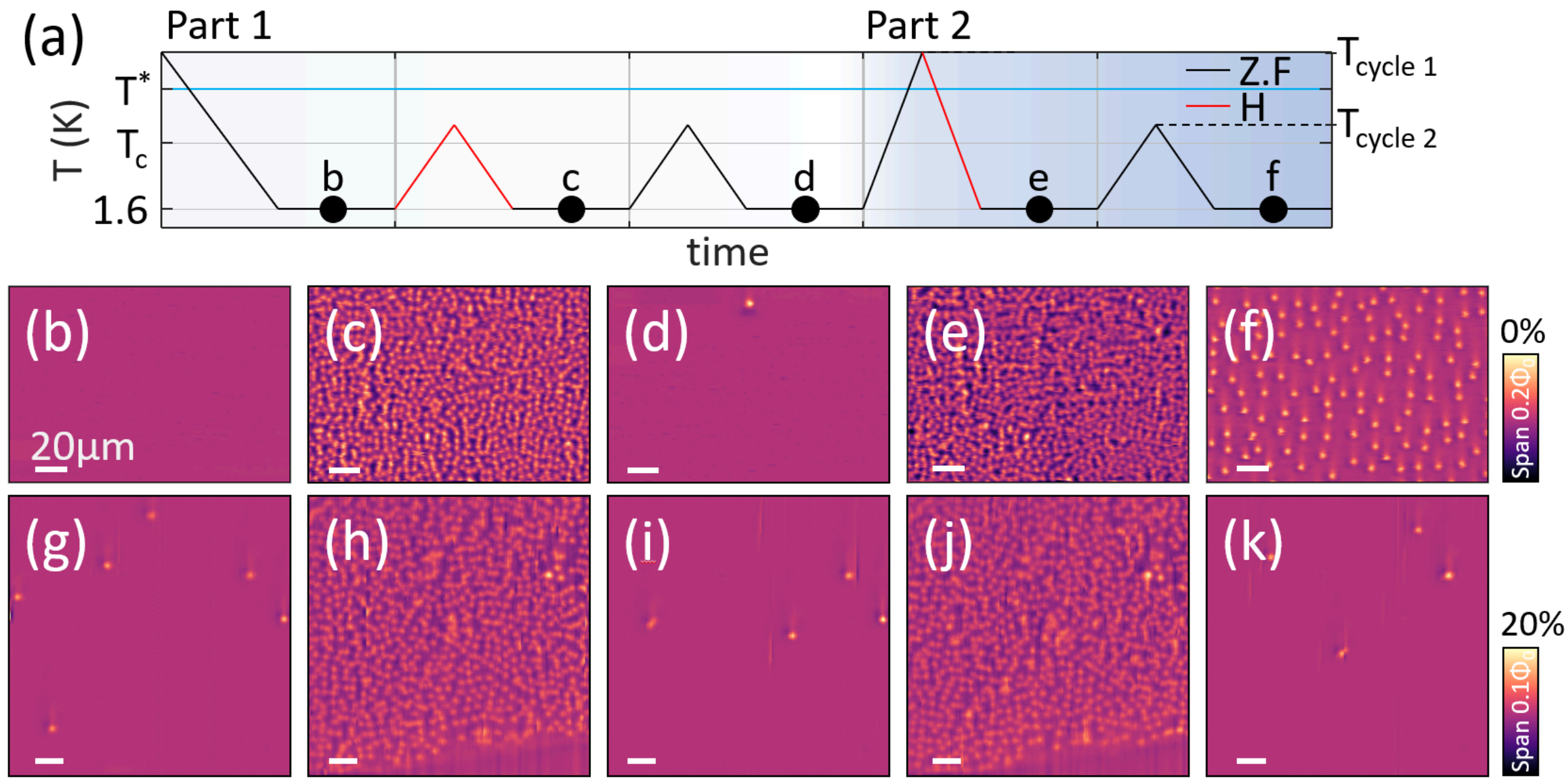

**Fig.1: Probing for memory in 4Hb-$TaS_2Se_0$ and 4Hb-$TaS_{1.8}Se_{0.2}$.** (a) Schematic of the experiment performed to determine the existence of spontaneous vortices. (b-f) Maps of the static magnetic landscape of 4Hb-$TaS_2Se_0$ with an area of 172*120 $\mu m^2$, taken at the points marked by the corresponding letters in (a), all at the same location, using H = 0.96G, $T_{cycle\ 1}$ = 4.6K, and $T_{cycle\ 2}$ = 3.1K. (b) No vortices are observed after zero-field cooling in the scanned area. (c) ~983 vortices are observed after field-cooling. (d) One vortex is observed after zero-field cooling. (e) ~1136 vortices are observed after the following field-cooling. (f) 121 spontaneous vortices are observed after zero-field cooling. (g-k) Magnetic maps of 4Hb-$TaS_{1.8}Se_{0.2}$ with an area of 195*194 $\mu m^2$, taken at the points marked by the corresponding letters in (a), using H = 0.69G, $T_{cycle\ 1}$ = 5K, $T_{cycle\ 2}$ = 3.1K. Panel f demonstrates that spontaneous vortices did not appear in this sample. All scale bars are 20 μm.

The memory-probing protocol we applied consists of two parts, as illustrated in Fig.1 (a). The first part tracks the behavior of the sample without "training", meaning we don't generate the memory effect. We expect to see trivial flux-trapping behavior in a superconductor. After the initial zero-field cooling through T*, the image shows almost no vortices, as expected. Then, we cooled down through Tc in the presence of a magnetic field and imaged the vortices again. The number of vortices reflects the cooldown field. A subsequent cycle through Tc with zero field returns to the expected density of almost no vortices, reflecting the zero field of our system. In the second part of the experiment, we generate the memory effect in the sample. We do this by cooling down through T* in the same magnetic field we used earlier, followed by zero-field cooling through Tc. If the sample exhibits memory, we observe a finite vortex density, even though it was cooled in zero field. These are the spontaneous vortices. Therefore, the simplest way to identify memory is to compare images 2d and 2f. If the first doesn't show vortices and the second shows spontaneous vortices, the sample exhibits magnetic memory.

Fig.1 (b-f) shows the testing results for the magnetic memory in a 4Hb crystal without any Se incorporated during growth. After cooling at our system's zero-field (which is ~3mG in our setup), almost no vortices are trapped (Fig.1 b, d), as expected for the dimensions of this scan area. However, after attempting to generate a memory (the second part of the protocol), we found 121 vortices in the samples. The sample behaves as if it was field-cooled in the presence of a field approximately 12.3% as strong as the training field used while field-cooling through T*. 4Hb without any selenium exhibits magnetic memory. The number of spontaneous vortices is proportional to the training field, allowing us to define the ratio as the memory strength, independent of the training field.

Observing memory in the sample without selenium is particularly interesting, as it clearly shows that the Se substitution is not responsible for this phenomenon. To accurately quantify the memory strength of the sample, we repeated the experiment several times to reduce the effect of the error that stems from the natural randomness of vortex distribution at low densities. These repetitions, using different fields and different locations on the sample, yielded an average memory strength of 13.7%.

The memory in the selenium-free sample was more substantial than in the 1% Se sample, originally detected as ~10% [8]. If this trend were to continue, increasing the selenium concentration should result in a reduction in memory strength. We repeated the experiment on a sample with 20% selenium. We did not find any observable memory in the sample, as shown in Fig.1 (g-k). The second part of the experiment (described above), where we train in a magnetic field through T* and attempt to delete the vortices by zero-field cooling around Tc, did not detect any spontaneous vortices.

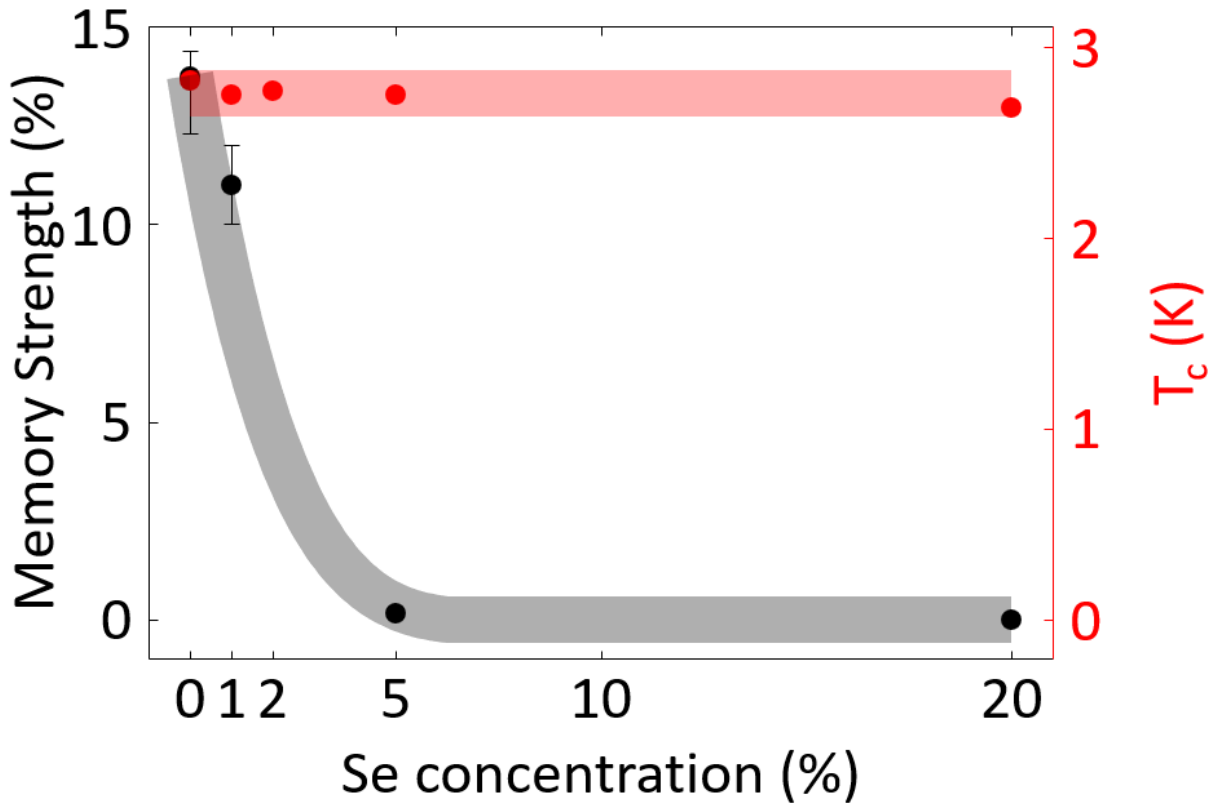

**Fig.2: Magnetic memory and Tc dependence on Se concentration in 4Hb-Ta$S_{2-x}Se_x$.** Memory strength (black) and Tc (red) as a function of the nominal Se concentration in 4Hb-Ta$S_{2-x}Se_x$. While the memory strength shows a gradual decrease with Se concentration, Tc remains relatively unchanged (For a detailed graph of Tc see Supplementary Fig.S2 **[29]**). In Supplementary Fig.S3 we show that 5% selenium still shows weak memory of about 0.15% **[29]**. Data of the 1% sample is taken from **[8]**. The curves are guides to the eye.

For the 5% Se sample, we observed a very weak magnetic memory of approximately 0.15% (see supplementary material Fig.S3 for details [29]). This marks the upper limit of Se concentration beyond which the system no longer exhibits detectable magnetic memory. While superconductivity remains unchanged, the disappearance of the memory with increasing Se concentration hints at the role of disorder in this process (Figure 2).

To examine the effect of Se concentration on T*, we train the system by cooling down in the presence of a magnetic field, verifying the existence of the memory, and then performing a sequence of zero-field cooling attempts, each to a slightly higher temperature. After each cycle, we image and count vortices, as shown in Fig.1 of Ref [8]. The memory is erased when the vortex density reflects the zero field of the system (almost no vortices for this typical image size). Fig.3 shows the results of this protocol on 0% Se, the memory was deleted at T* = 3.6 K, which is consistent with the results of the 1% samples [8]. At 5%, T* is around 3.5-4K. The large error bar is a result of the weak memory level, preventing an accurate determination of T*, Supplementary Fig.S4 [29].

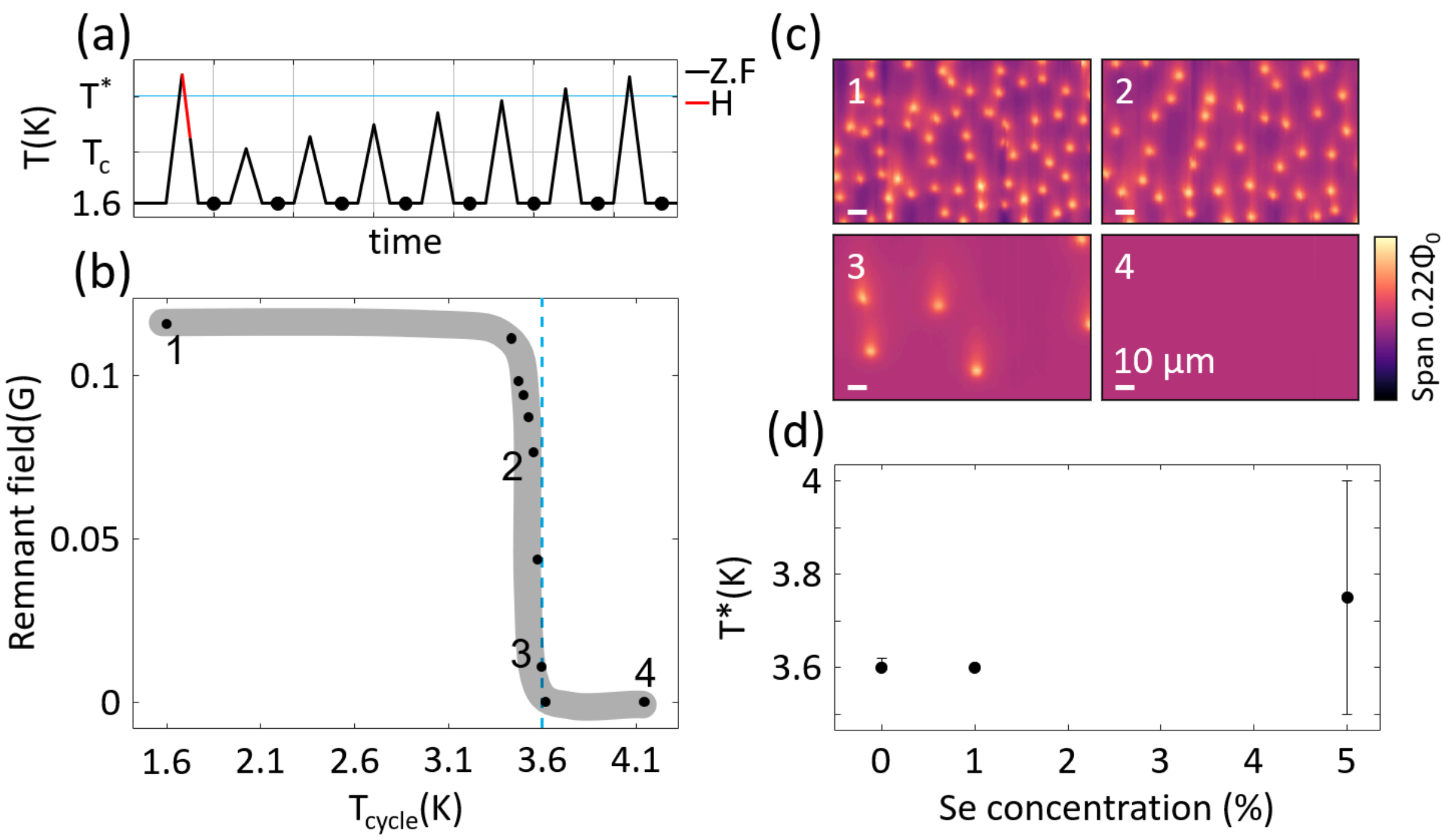

**Fig.3: Memory survives up to T* in 0% Se.** (a) Schematic of the experimental procedure for measuring what is T* in a sample, with H = 0.8G. The black lines represent times when the zero field was activated, while the red lines represent times H was activated; black circles mark scans. (b) The calculated remnant field, plotted as a function of the cycling temperature used during the zero-field cooling cycle preceding each scan. Remnant flux is calculated by multiplying the vortex count with the flux quantum $\Phi_0$ (~20.6 G*µm$^2$) and dividing that by the scan area (c) Magnetometry maps of the remnant vortices, following zero-field cooling, as seen in (a). (d) T* as a function of the nominal Se concentration in 4Hb-$TaS_{2-x}Se_x$. The large uncertainty at 5% Se arises from the weak memory signal, which resulted in an insufficient vortex density for a reliable determination of T*. See Supplementary Fig.S4 for details**[29]**. The 20% Se sample is not included because no memory signal was detected.

The effect of Se substitution on the magnetic memory phenomenon can be disentangled by analyzing two distinct measures: (1) the characteristic temperature T*, at which memory is lost, and (2) the strength of the memory effect, defined as the ratio between the spontaneous magnetic field and the initial training field. Our results reveal a difference in how these two quantities respond to increasing Se concentration. The temperature T* remains essentially unchanged, indicating a robust onset temperature for the memory effect. In contrast, the memory strength is highly sensitive to Se concentration, changing significantly from 14% to 0.15% over the same range.

The contrasting responses of T* and the memory strength to Se substitution are consistent with a two-stage scenario for the emergence of magnetic memory. In this picture, a weak and possibly undetectable magnetic response develops below T*, marking the onset of the hidden state. Upon entering the superconducting phase, this underlying response can be amplified or converted into observable spontaneous vortices through the nucleation of quantized flux lines. The observation that T* remains essentially unchanged with Se substitution, while the vortex density and memory strength are strongly suppressed, suggests that Se substitution primarily affects the process by which the hidden state manifests itself in the superconducting phase, rather than the temperature scale at which the hidden state emerges. Thus, whatever microscopic changes are introduced by Se substitution appear to have a much stronger effect on the generation or stabilization of spontaneous vortices than on the underlying hidden-memory state itself.

An intriguing aspect of this behavior is that the superconducting transition temperature Tc remains nearly unchanged with increased Se concentration, even as the memory strength diminishes significantly. This raises a key question: what property of the superconductor is being altered by Se substitution, affecting its ability to amplify the underlying remnant magnetic field? Notably, this effect cannot be attributed to enhanced vortex pinning, which typically becomes stronger with increasing disorder and would instead favor the retention of vortices.

One possible explanation is that Se substitution modifies the intrinsic inhomogeneous electronic landscape of 4Hb-$TaS_2$ rather than simply increasing a homogeneous disorder scattering rate. Recent STM/STS measurements on the same sample series have shown that the incommensurability between the 1T and 1H layers produces spatial variations in interlayer distance, hybridization, and charge transfer, which are present even in nominally Se-free samples [14]. Se substitution further modifies the interlayer separation and therefore provides a natural means of perturbing this landscape. Such changes may influence the coupling between the CDW, correlated 1T states, and superconductivity without strongly affecting the superconducting transition temperature. In addition, previous measurements have shown that the CDW order in the 1H layers is rapidly suppressed with increasing Se concentration [17]. It is therefore possible that the strong reduction of the spontaneous-vortex response reflects a sensitivity to the spatial electronic and CDW landscape rather than a simple suppression of superconductivity. In this picture, Tc can remain nearly unchanged while the ability of the superconducting state to convert the underlying hidden magnetic response into stable spontaneous vortices is strongly modified.

The microscopic origin of the magnetic memory effect in 4Hb-$TaS_2$ remains an open question, with several theoretical proposals invoking different forms of correlated magnetism and coupling to superconductivity [9–13]. A common feature of several scenarios is that an underlying response, which may be too weak to detect directly above Tc, becomes observable in the superconducting state through the generation of spontaneous vortices. Our results provide an important constraint on the role of disorder in this process. In particular, the observation of a strong memory effect in nominally Se-free 4Hb-$TaS_2$ demonstrates that Se substitution is not required to generate the memory. Moreover, increasing Se concentration strongly suppresses the spontaneous-vortex response while leaving superconductivity largely intact. This behavior disfavors scenarios in which Se-induced disorder, impurity pinning, or disorder-stabilized magnetic clusters are essential for forming or enhancing the memory, while remaining compatible with mechanisms in which an intrinsic hidden state is manifested through its coupling to superconductivity.

In this work, we investigated how dilute Se substitution affects superconductivity and hidden magnetic memory in 4Hb-$TaS_{2-x}Se_x$. We find that the memory exists in nominally Se-free crystals, demonstrating that Se is not required for the phenomenon, while its spontaneous-vortex manifestation is rapidly suppressed with increasing Se concentration. By contrast, both the residual-resistance ratio and Tc change only weakly across the substitution series, and $T^*$ remains approximately unchanged where it can be determined. These contrasting responses indicate that Se substitution does not simply tune the system through a conventional increase in transport scattering, but instead modifies a more microscopic structural or electronic landscape to which the magnetic-memory response is particularly sensitive. One possible origin is a modification of the intrinsic incommensurate and CDW landscape, which may affect the coupling between the hidden state and superconductivity without strongly altering Tc. More broadly, 4Hb-$TaS_2$ provides a model system in which robust unconventional superconductivity coexists with a strongly scattering normal state, while an intertwined magnetic phenomenon is extraordinarily sensitive to a subtle local perturbation.

**Data availability**: The data supporting the findings of this study are available from the authors upon reasonable request.

**Acknowledgements:** We thank Erez Berg, Jonathan Ruhman, Avi Klein and Roni Gofman for fruitful discussions. This work was supported by the German–Israeli Project Cooperation (DIP), grant no. 3970/1-1. The work at Bar-Ilan was supported by the European Research Council Grant No. ERC-866236, the Israeli Science Foundation (ISF) Grant No. ISF-228/22. The work at the Technion was supported by the ISF Grant No. 568/25.

**Author contributions**: S.G. and B.K. performed the scanning SQUID measurements and carried out the data analysis. S.K.P., S.P., I.F., I.M, N.R and A.K. grew the samples and performed the transport measurements. S.G., B.K., and A.K. discussed the results and wrote the manuscript.

# Supplemental Material

Shalev Gur [1], Sajilesh K.P.[2], Shubam Purwar[2], Irena Feldman [2], Ilay Mangel[2], Nitzan Ragoler[2], Amit Kanigel [2*] and Beena Kalisky [1*]

1. Bar Ilan University, Ramat Gan, Israel
2. Technion, Haifa, Israel

*amitk@physics.technion.ac.il, *beena@biu.ac.il

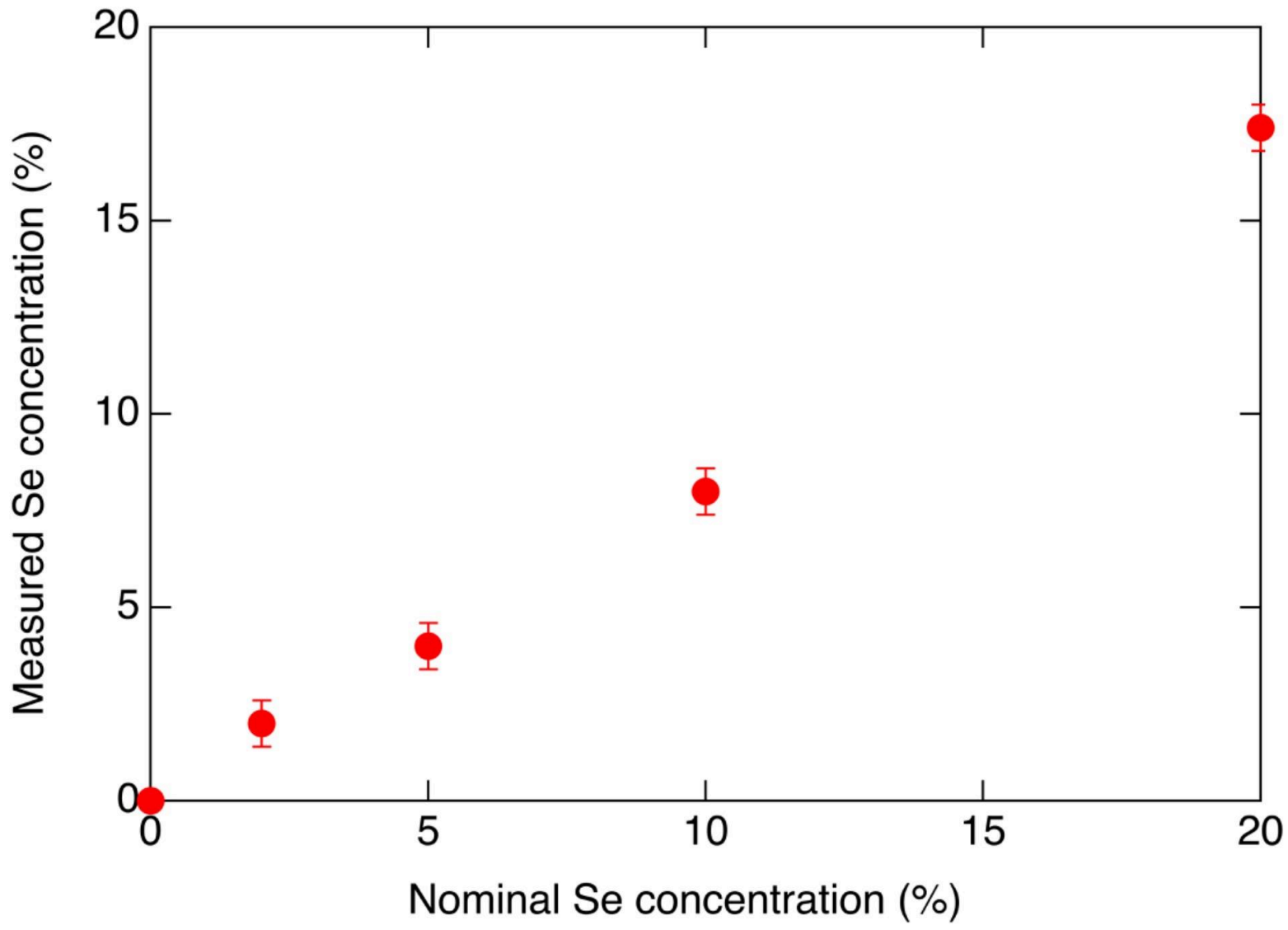

**Fig.S1: Selenium concentration.** The amount of Se in the crystals versus the amount of Se in the starting mixture, measured by X-ray diffraction and electron energy dispersive spectroscopy in a scanning electron microscope.

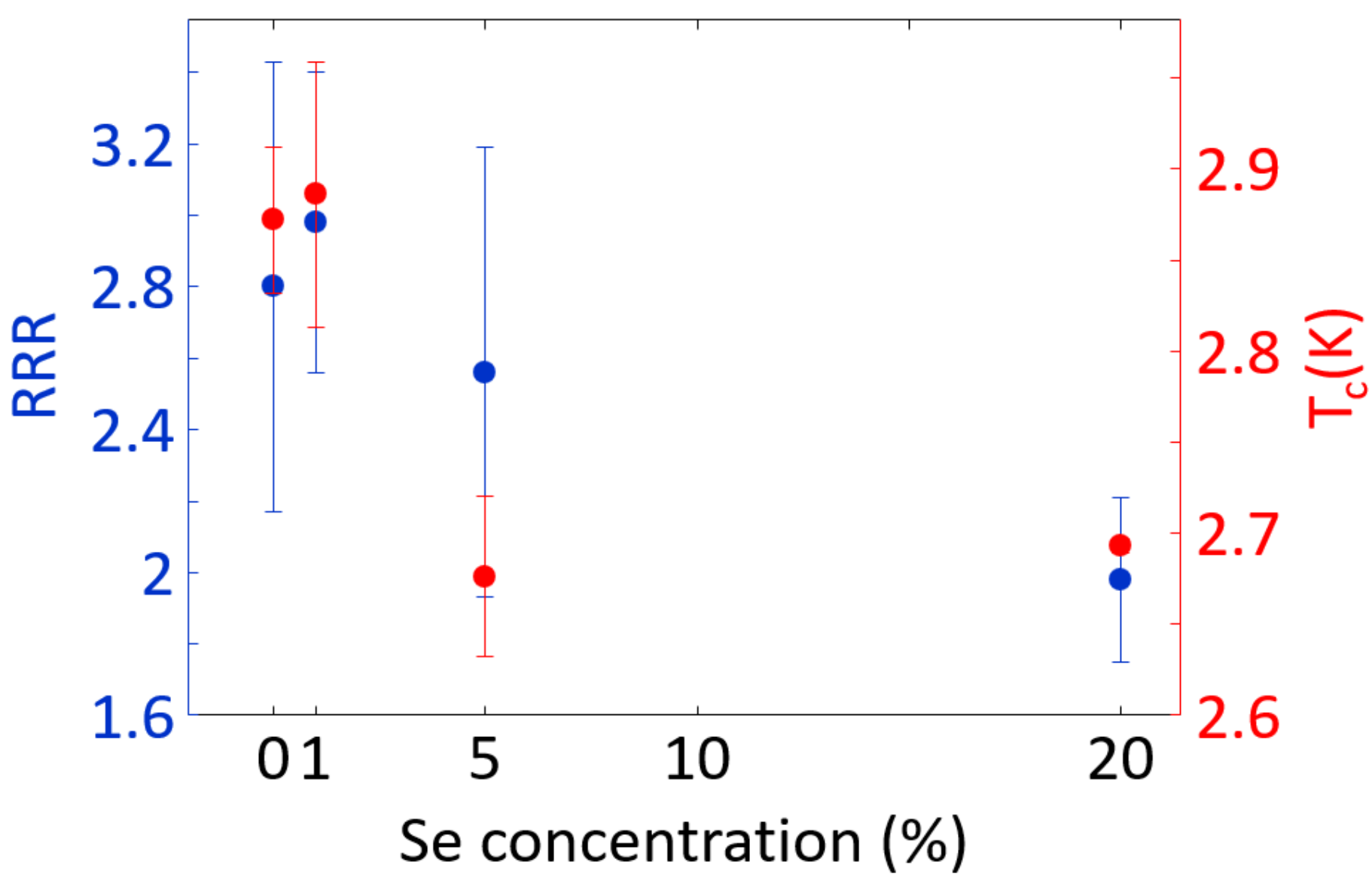

**Fig.S2: RRR and $T_c$ as a function of Se concentration.** RRR (blue) and Tc (red) as a function of the nominal Se concentration in 4Hb-$TaS_{2-x}Se_x$. While Tc shows a moderate reduction of ~0.2K in between Se concentrations of 1% and 5%, RRR remains comparable throughout all samples. RRR and Tc are averaged on 5 different crystals from each concentration, from several growth batches. The error bar is the standard deviation.

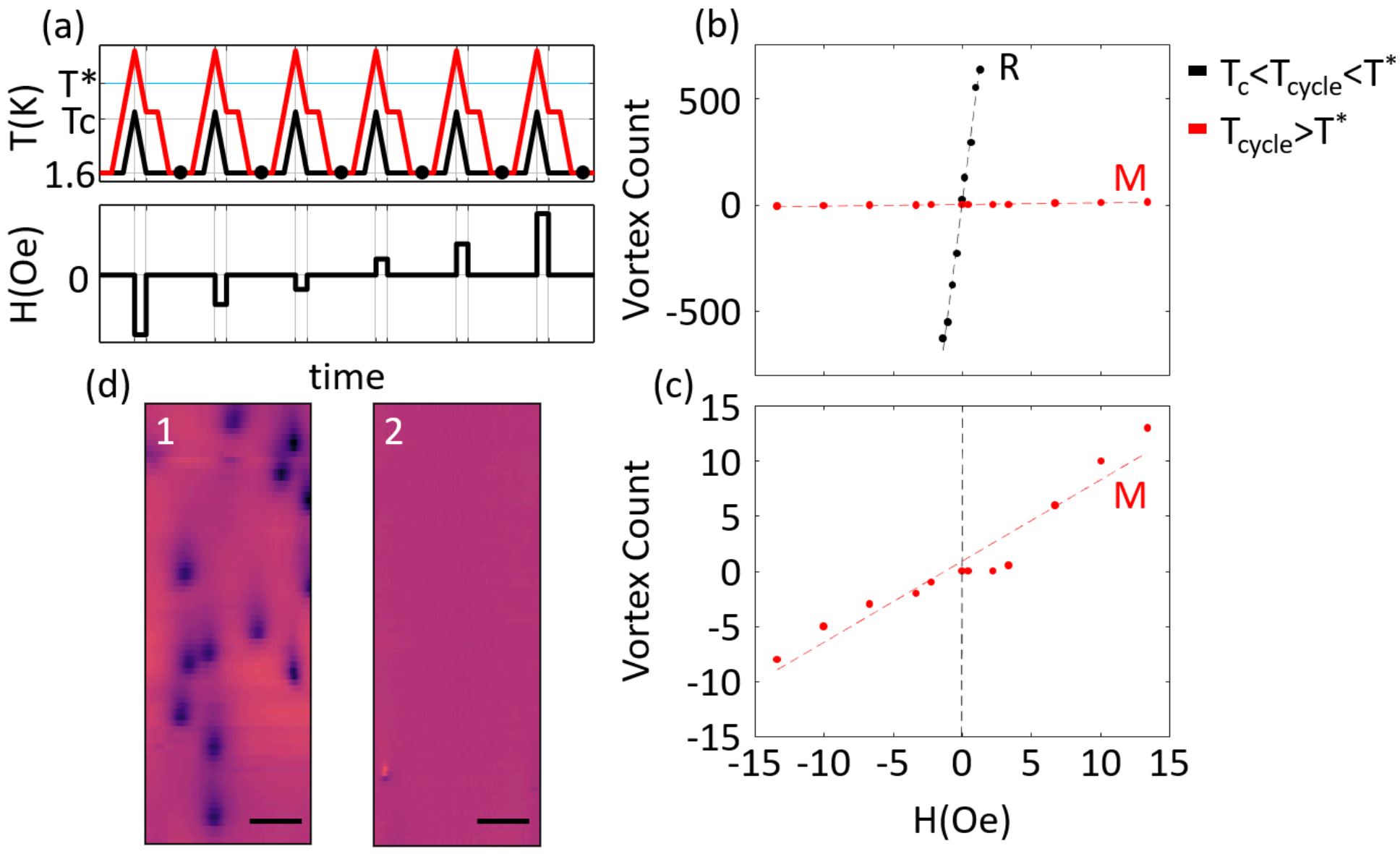

**Fig.S3: Memory strength in 4Hb-$TaS_{1.95}Se_{0.05}$.** (a) The experimental procedure we used for probing the hidden magnetic memory. The red cycles generate a magnetic landscape trapped with only spontaneous vortices using $T_{cycle}$ = 6K, we call this data set M (for memory). In comparison, the black cycles only trap “regular” vortices induced by field-cooling an untrained sample, using $T_{cycle}$ = 2.85K, we call this data set R (for regular). (b) The vortex count of regular (black) and spontaneous (red) vortices, plotted against the field used for that specific cycle. The fields used to gather R are in the range of ±1.33G, while the fields used to gather M are in the range of ±13.3G. There are two ways to calculate the memory strength with this data. The first step is to obtain a linear fit for both data sets and then divide the inclines. The second is by dividing M/R for each point, extrapolating R for the entire field range. Following the first method yields memory strength of 0.146%. The second one yields a memory strength vector with a mean value of 0.14%, and STD of 0.0573%. (c) The same graph as in (b), focused on the memory-related data. (d) Magnetometry maps obtained in the fifth and third states of the memory probing protocol, as described in Fig.1 (a) in the main text. Scale bar is 20 µm. (1) The fifth state corresponds to a zero-field cooled state that followed a field-cooling through T*, with H = 13.3G. This training protocol yielded 13 spontaneous vortices. (2) The third state corresponds to a zero-field cooled measurement that followed a field-cooling from 2.85K. This shows that an untrained sample does not show spontaneous vortices.

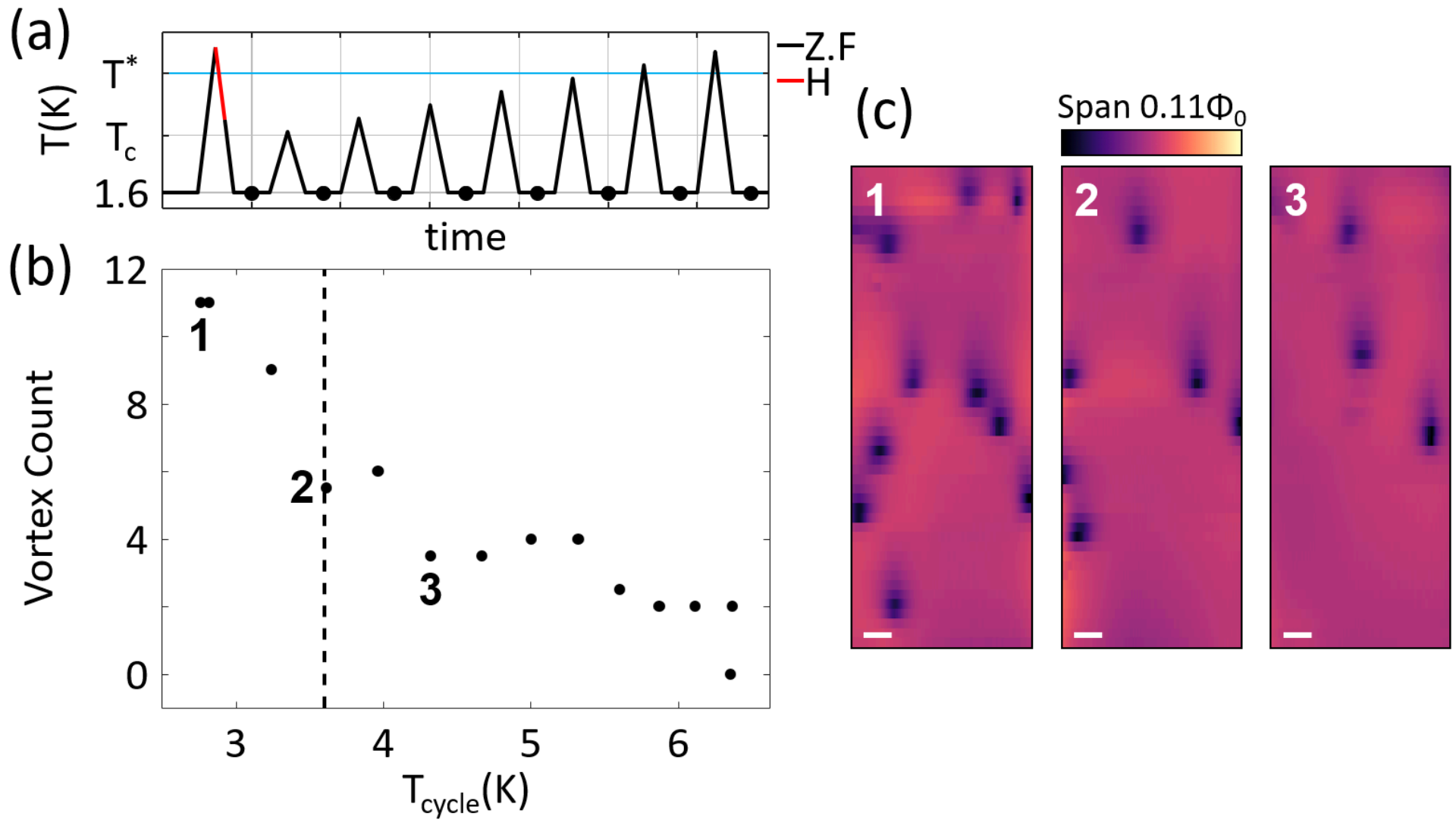

**Fig.S4: Finding T* in 4Hb-$TaS_{1.95}Se_{0.05}$.** (a) The experimental procedure we used for finding T*, as seen in the main text, H = 13.3G, black circles mark scans. (b) The calculated remnant field, plotted as a function of the cycling temperature used in the zero-field cooling cycle done prior to each scan. (c) Magnetometry maps of the remnant vortices, following zero-field cooling, as seen in (a).